\documentclass[aps,prx,twocolumn,showpacs]{revtex4-1}
\usepackage{amsfonts}
\usepackage{amsmath}
\usepackage{amssymb}
\usepackage{graphicx}
\usepackage{txfonts}
\usepackage{xcolor}
\usepackage[T1]{fontenc}%
\DeclareMathOperator{\tr}{tr}
\DeclareMathOperator{\Det}{Det}

\begin{document}

\title{Genuine tripartite entanglement as a probe of quantum phase transitions in a spin-1 Heisenberg chain with single-ion anisotropy}
\author{Chon-Fai Kam}
\email{Email: dubussygauss@gmail.com}
\affiliation{Department of Mathematics, Faculty of Science and Technology, University of Macau, Avenida da Universidade, Taipa, Macau, People's Republic of China}
\author{Yang Chen}
\email{Email: yangbrookchen@yahoo.co.uk}
\affiliation{Department of Mathematics, Faculty of Science and Technology, University of Macau, Avenida da Universidade, Taipa, Macau, People's Republic of China}

\begin{abstract}
We study the quantum phase transitions of spin-1 Heisenberg chains with an easy-axis anisotropy $\Delta$ and a uniaxial single-ion anisotropy $D$ using a multipartite entanglement approach. The genuine tripartite entanglement between the spin blocks, measured by the tripartite qutrit hyperdeterminant, is calculated within the quantum renormalization group method. Using this approach, the phase boundaries between the topological Haldane, large-D and anti-ferromagnetic N\'eel phases are determined in the half $\Delta-D$ plane with $\Delta>0$. When the size of the spin blocks increases, the genuine tripartite entanglement between the blocks exhibits a nonzero plateau in the topological Haldane phase, and experiences abrupt drops at both the phase boundaries between the Haldane--large-D and Haldane--N\'eel phases, which justifies the usage of genuine multipartite entanglement as a probe of topological phases in spin systems.  
\end{abstract}


\maketitle
\section{Introduction}
Quantum phase transition (QPT) is a transition between different quantum phases accompanied by a qualitative change in the ground state of many-body correlated systems driven by a small variation in the external fields \cite{sachdev2011quantum}. Traditionally, quantum phase transitions are characterized by spontaneous symmetry breaking in the Landau paradigm of phase transitions, which is based on the concept of local order parameters \cite{landau1937on}. A well known example of the traditional phase transition is the  ferromagnet–paramagnet transition in spin systems, of which the magnetization serves as a local order parameter for the magnetic order. Over the last few decades, some striking examples of quantum phases which falls beyond the symmetry description have been discovered \cite{wen2017colloquium}. Such phases, known as the topological phases, cannot be characterized by local order parameters and long-range correlations used in the Landau symmetry breaking theory \cite{chen2010local}. A first example of topological phases is the Haldane phase of odd-integer spin chain \cite{haldane1983nonlinear, haldane1983continuum}, which is characterized by a double degeneracy of the entanglement spectrum \cite{pollmann2010entanglement}.

In recent years, there has been increasing interest in utilizing concepts and ideas from quantum information to explore quantum phase transitions \cite{zeng2019quantum}. Along the quantum information approach, one of the most popular directions is the ground state fidelity and the associated fidelity susceptibility \cite{amico2008entanglement, gu2010fidelity, braun2018quantum}, which is the overlap intensity between two many-body ground states corresponding to the Hamiltonians differing by a small perturbation \cite{rossini2018ground}. A qualitative change in a many-body ground state at the quantum phase transition point is characterized by an abrupt drop in the ground-state fidelity, accompanied by a sharp peak in the associated fidelity susceptibility, irrespective of the existence of local order parameter  \cite{gu2010fidelity}. This particular property of the ground-state fidelity makes it possible to identify topological phase transitions in many-body systems. Besides the ground-state fidelity and the fidelity susceptibility, there are other approaches to characterize topological phase transitions, such as the entanglement entropy in the ground-state of many-body systems \cite{kitaev2006topological, levin2006detecting, cho2017quantum, maslowski2020quasiperiodic}, and the quantum discord of the ground state which is the difference between the quantum analogues of two classically equivalent expressions of mutual information \cite{shan2014scaling, maziero2010quantum, sarandy2009classical, dillenschneider2008quantum, lin2021single}.

Unlike the entanglement entropy and the quantum discord, which are essentially bipartite entanglement or correlations, multipartite entanglement in many-body systems is relatively less explored. Tripartite entanglement has been considered in the spin-1/2 Heisenberg and XY models \cite{guhne2005multipartite}, and the spin-1/2 XXZ chains in a transverse magnetic field \cite{bruss2005multipartite}. Multipartite entanglement in two-dimensional topological systems such as the toric code has been analyzed by using the geometric entanglement of blocks \cite{orus2014geometric, orus2014geometric}. Multipartite entanglement is a unique resource to quantum information processing \cite{wootters1998quantum}, which cannot be increased by local operations performing on spatially separated systems \cite{horodecki2009quantum}. As multipartite entangled states are classified and quantified by multipartite entanglement measures which are invariant under local operations, a question naturally arises: can multipartite entanglement measures be able to detect topological phases in many-body systems?

In this work, we will study the relations between genuine tripartite entanglement and quantum phase transitions in a spin-1 Heisenberg XXZ chain with single-ion anisotropy by using the quantum renormalization group (QRG) method. We will show that after the real space renormalization procedures in which clusters of neighboring sites are merged into a large spin block, the genuine tripartite entanglement between the blocks, i.e., the tripartite qutrit hyperdeterminant, shows a nonzero plateau in the topological Haldane phase, and exhibits an abrupt drop at the phase boundaries between the Haldane--large-D phases, and the Haldane--N\'eel phases.

The organization of the paper is as follows. In Sec.\:\ref{II}, we discuss the spin-1 Heisenberg XXZ model with single-ion anisotropy, the properties of its different quantum phases, and the quantum renormalization group method for spin chains. In Sec.\:\ref{III}, we explicitly evaluate the genuine tripartite entanglement between the spin blocks by using the hyperdeterminant, which can be constructed by the three fundamental SL$(3,\mathbb{C})$ invariants of degree 6, 9 and 12 respectively. In Sec.\:\ref{IV}, we show that the genuine tripartite entanglement between the spin blocks, measured by the hyperdeterminant, can be utilized as a probe of the topological Haldane phase in the spin-1 Heisenberg chain. In Se.\:\ref{V}, we conclude our study and outline some future opportunities.

\section{The Spin model and the Quantum Renormalization Group}\label{II}
To better illustrate our quantum information and multipartite entanglement approach to quantum phase transitions, we begin with a spin-1 Heisenberg XXZ anti-ferromagnet with a uniaxial single-ion anisotropy term, which is described by the Hamiltonian  \cite{langari2013ground}
\begin{equation}\label{OriginalHamiltonian}
    H=J\sum_{i=1}^N[S_i^xS_{i+1}^x+S_i^yS_{i+1}^y+\Delta S_i^zS_{i+1}^z+D(S_i^z)^2],
\end{equation}
where $S_i^{\alpha}$ ($i=1,2,3$) are the spin-1 operators at the $i$-th lattice site, $J>0$ is the anti-ferromagnetic exchange coupling, $\Delta$ quantifies the easy-axis anisotropy, and $D$ is the uniaxial single-ion anisotropy parameter, which can be experimentally adjusted by nuclear electric resonance \cite{asaad2020coherent}. One may set $J=1$ to fix the energy scale. The full phase diagram for the spin-1 Heisenberg XXZ chain with a uniaxial single-ion anisotropic term was extensively studied \cite{botet1983ground, schulz1986phase, kitazawa1996phase, chen2003ground, venuti2007quantum, tzeng2008fidelity, hu2011accurate}, which consists of six different phases, i.e., the Haldane phase, the large-D phase, two XY phases, the ferromagnetic phase, and the N\'eel phase \cite{chen2003ground}. 

Both the ferromagnetic and N\'eel phases have magnetic order. In the ferromagnetic phase, all the spins point in the same direction, which are characterized by the non-zero spin-spin correlations $\langle S_i^zS_{i+n}^z\rangle$, where $\langle\cdots\rangle$ represents the expectation value in the ground state; while in the N\'eel phase, all the spins at the nearest neighbor sites are aligned in the opposite directions, which are characterized by the non-zero N\'eel order parameter $(-1)^n\langle S_i^zS_{i+n}^z\rangle$ \cite{hatsugai1991numerical}. In the XY phases, the spins prefer to lie in the $xy$-plane  rather than along $z$-direction \cite{lee2020bell}. The XY phases are gapless, which leads to a power-law decaying spin-spin correlations $\langle S_i^+S_{i+r}^-\rangle$ and $\langle (S_i^+)^2(S_{i+r}^-)^2\rangle$ in the XY1 and XY2 phases respectively \cite{schulz1986phase, alcaraz1992critical, kitazawa1996phase}. Both the Haldane and large-D phases are gapful, which leads to an exponentially decaying spin–spin correlations \cite{haldane1983nonlinear, haldane1983continuum}. But in the Haldane phase, although the N\'eel order parameter vanishes as the ground state is disordered, there is a hidden order in the Haldane phase characterized by de Nijs and Rommelse's nonlocal topological string order parameter $-\langle S_i^z\exp(i\pi\sum_{k=i+1}^{j-1}S_k^z)S_j^z\rangle$ \cite{den1989preroughening, ren2020quantum}, when the system preserves the hidden $Z_2\times Z_2$ symmetry \cite{lee2020bell,kennedy1992hidden}, corresponding to the rotation by $\pi$ around the $z$ and $x$ axes respectively \cite{hatsugai1991numerical}.

Between these phases, there are several types of quantum phase transitions, such as a Gaussian topological phase transition between the gapful Haldane and large-D phases with opposite parity symmetry \cite{langari2013ground, chen2003ground}, an infinite-order Berezinskii-Kosterlitz-Thouless (BKT) transition between the gapless XY phases and the gapful Haldane or large-D phases \cite{chen2003ground}, an Ising transition between the N\'eel and Haldane phases \cite{chen2003ground}, and a first-order transition between the ferromagnetic phase and the large-D or XY phases \cite{chen2003ground}.

\begin{figure}[tbp]
\includegraphics[width=0.8\columnwidth]{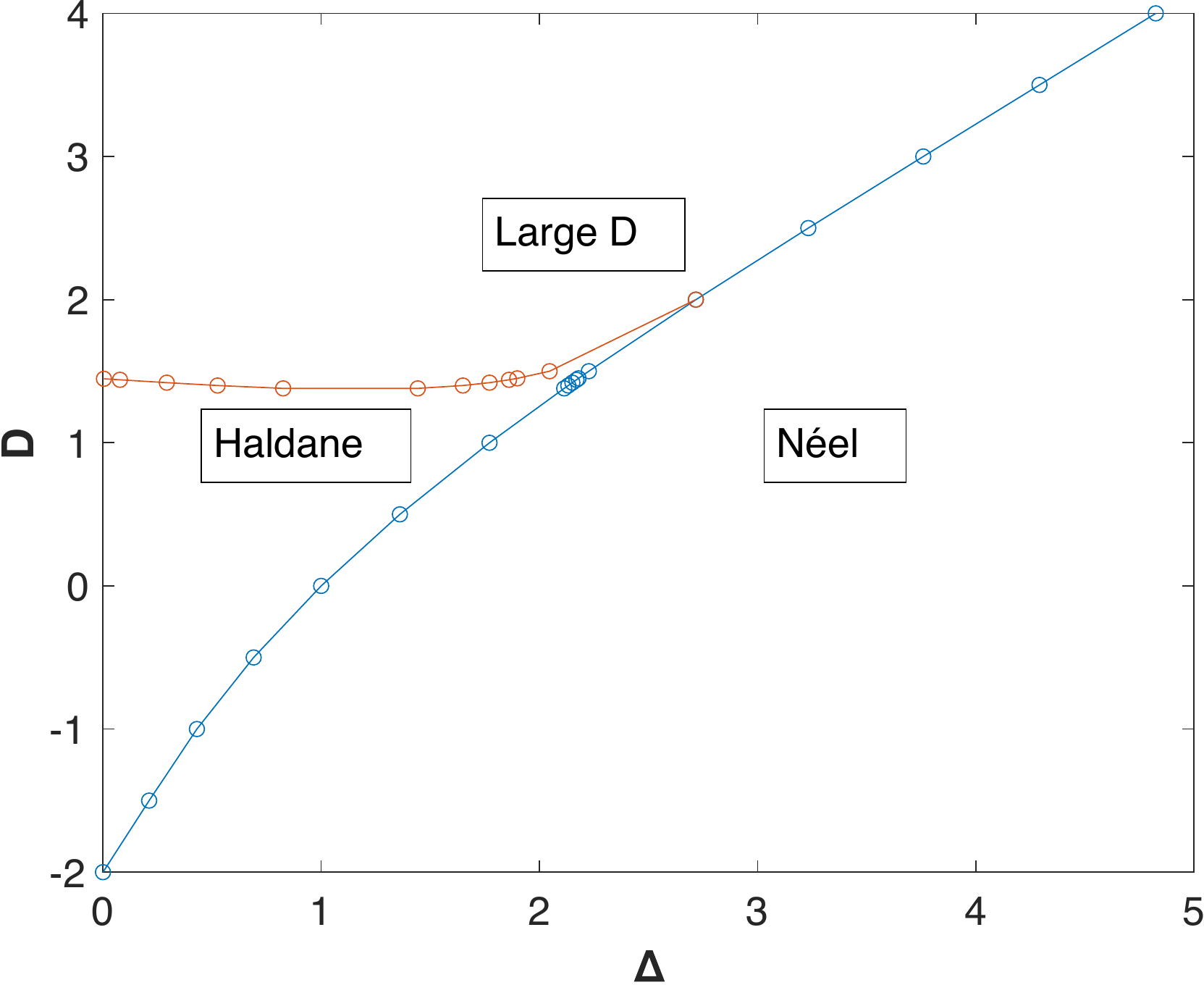}
\caption{Schematic of the phase diagram for a spin-1 Heisenberg XXZ antiferromagnet with single-ion anisotropy in the $\Delta-D$ plane. Here, we consider only a non-negative easy-axis anisotropy, i.e., $\Delta\geq 0$. For a small single-ion anisotropy parameter with $-2\lessapprox D\lessapprox 1.38$, there are two phases, i.e., the Haldane phase and the antiferromagnetic Néel phase. By contrast, for a small single-ion anisotropy parameter with $1.38 \lessapprox D\lessapprox 2$, there exist three phases, i.e., the Haldane phase, the Large D phase and the antiferromagnetic Néel phase. For a large positive single-ion anisotropy parameter, $D\gg 2$, there are another two phases, i.e., the Large D phase and the antiferromagnetic Néel phase. Finally, for a large negative single-ion anisotropy, $D\ll -2$, only the antiferromagnetic Néel phase exists.}
\label{HeisenbergXXZPhaseDiagram}
\end{figure}

Here we assume the system to be antiferromagnetic, i.e., $\Delta>0$. On the half-plane with $\Delta>0$, there are three distinct phases: a Haldane phase, a large-D phase, and a N\'eel phase \cite{langari2013ground}. All these phases have a finite energy gap above the ground state \cite{tzeng2008scaling}. The phase diagram of spin-1 Heisenberg XXZ with a uniaxial single-ion anisotropy is shown in Fig.\:\ref{HeisenbergXXZPhaseDiagram} by use of the quantum renormalization group (QRG) method described below. For a large negative single-ion anisotropy, i.e., $D\ll-2$, the system prefers the $|\uparrow\rangle$ and $|\downarrow\rangle$ states rather than the $|0\rangle$ state. Hence, one expects that the N\'eel order parameter is non-zero, which corresponds to the N\'eel phase \cite{hatsugai1991numerical}. On the other hand, for a large positive single-ion anisotropy, i.e., $D\gg 2$, the system prefers the $|0\rangle$ state in the large-D phase, and thus all the spin-spin correlations are expected to be short ranged \cite{hatsugai1991numerical}. 

The quantum renormalization group (QRG) method is a real-space renormalization group method based on Kadanoff's block-spin transformations \cite{langari2013ground}. To carry out the QRG procedure, one needs to first decompose the lattice into a collection of spin blocks, so that the Hamiltonian can be written as a sum of the block Hamiltonians and the inter-block interactions, i.e., $H=H^B+H^{BB}$. Then one needs to diagonalize the block Hamiltonian to find the low-lying energy eigenstates of each block, and construct a \textit{truncation operator} $T^\dagger:\mathcal{H}\rightarrow\mathcal{H}^\prime$ (or equivalently an \textit{embedding operator} $T$: $\mathcal{H}^\prime\rightarrow \mathcal{H}$) which maps the most important subspace of the original Hilbert space to the renormalized Hilbert space \cite{martin1996analytic}. To ensure that the renormalized Hamiltonian $H^\prime$ and the original Hamiltonian $H$ have a common low-lying energy spectrum, one needs to impose the condition $HT=TH^\prime$ on $H^\prime$. It implies that if $|\psi^\prime\rangle$ is an eigenstate of the renormalized Hamiltonian $H^\prime$ with an eigenvalue $E^\prime$, $T|\psi^\prime\rangle$ is an eigenstate of the original Hamiltonian $H$ with the same eigenvalue, i.e., $HT|\psi^\prime\rangle=TH^\prime|\psi^\prime\rangle=E^\prime T|\psi^\prime\rangle$ \cite{martin1996analytic}. Moreover, to ensure that $|\psi\rangle=T|\psi^\prime\rangle\rightarrow |\psi^\prime\rangle=T^\dagger|\psi\rangle$, one needs to impose the conditions $T^\dagger T=I_{\mathcal{H}^\prime}$ and $TT^\dagger \neq I_{\mathcal{H}}$ on the truncation and embedding operators, where the last condition ensures that the two Hilbert spaces $\mathcal{H}$ and $\mathcal{H}^\prime$ are not isomorphic. Hence, one may construct the renormalized Hamiltonian $H^\prime$ as \cite{langari2013ground}
\begin{equation}
    H^\prime = T^\dagger H T.
\end{equation}
It implies that the mean energy of the renormalized Hamiltonian with respect to its eigenstate coincides with that of the original Hamiltonian with respect to the state obtained through the embedding operator: $\langle\psi^\prime|H^\prime|\psi^\prime\rangle=\langle \psi|H|\psi\rangle$. In other words, $|\psi\rangle=T|\psi^\prime\rangle$ can be used as a variational state for the eigenstates of the original Hamiltonian \cite{martin1996analytic}. Finally, the renormalized Hamiltonian $H^\prime$ has to be self-similar to the original Hamiltonian, so that one can determine the renormalization group flow of the coupling constants \cite{langari2013ground}.

\begin{figure}[tbp]
\includegraphics[width=0.8\columnwidth]{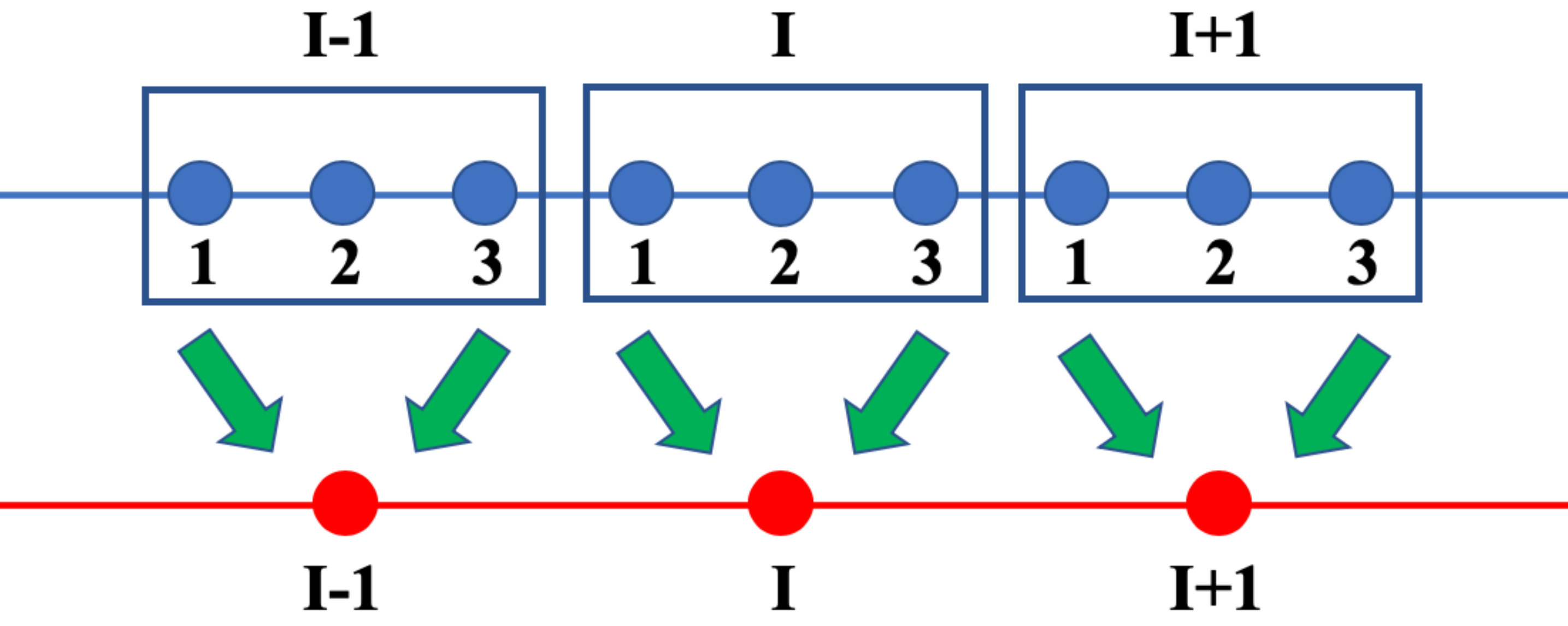}
\caption{Schematic of the quantum renormalization group (QRG) procedure. The spin-1 chain is decomposed into a collection of spin blocks, and each spin block is represented by an effective spin (red dot) after the renormalization process.}
\label{RGFigure}
\end{figure}

For our present case, we decompose the original Hamiltonian Eq.\:\eqref{OriginalHamiltonian} into a collection of spin blocks, each of which contains three spins (see Fig.\:\ref{RGFigure}), where $H^B\equiv \sum_{I=1}^{N/3}h_I^B$, $H^{BB}\equiv \sum_{I=1}^{N/3}h_{I,I+1}^{BB}$, and $h_I^B$ and $h_{I,I+1}^{BB}$ are determined by \cite{langari2013ground}
\begin{align}\label{BlockSpin}
    h_I^B&=\sum_{i=1}^2(S_{I,i}^xS_{I,i+1}^x+S_{I,i}^yS_{I,i+1}^y+\Delta S_{I,i}^zS_{I,i+1}^z)+D\sum_{i=1}^3(S_{I,i}^z)^2,\nonumber\\
    h_{I,I+1}^{BB}&=S_{I,3}^xS_{I+1,1}^x+S_{I,3}^yS_{I+1,1}^y+\Delta S_{I,3}^zS_{I+1,1}^z,
\end{align}
where $S_{I,i}^\alpha$ denotes the $\alpha$-component of the $i$-th spin in the $I$-th spin block. The energy eigenstates of $h_I^B$ can be calculated exactly, and the three lowest eigenstates of are denoted by $|\psi_0\rangle$ and $|\psi_\pm\rangle$ which obey $h_I^B|\psi_0\rangle=E_0|\psi_0\rangle$ and $h_I^B|\psi_\pm\rangle=E_1|\psi_\pm\rangle$ respectively (see Appendix \ref{ExpressionHp}). Hence, one may construct the embedding operator for each individual spin block as \cite{langari2013ground}
\begin{equation}
    T_I\equiv|\psi_+\rangle\langle +1|+|\psi_0\rangle\langle 0|+|\psi_-\rangle\langle -1|,
\end{equation}
where the set of states $|\pm\rangle$ and $|0\rangle$ form a base for the renormalized Hilbert space of each spin block. Hence, the global embedding operator $T:\mathcal{H}^\prime\rightarrow\mathcal{H}$ can be directly constructed by $T=\otimes_{I=1}^{N/3}T_I$. The renormalized coupling constants are determined from the original ones by \cite{langari2013ground}
\begin{equation}\label{RGflow}
    J^\prime = X_{ren}^2J,
    \Delta^\prime = \frac{Z_{ren}^2}{X_{ren}^2}\Delta,
    D^\prime = \frac{\epsilon_1-\epsilon_0}{X_{ren}^2},
\end{equation}
where $\epsilon_0$ is the smallest root of the cubic equation $\epsilon_0^3+(\Delta-4D)\epsilon_0^2+(4D^2-2D\Delta-6)\epsilon_0+8D=0$, $\epsilon_1$ is the smallest root of the quartic equation $\epsilon_1^4+2(\Delta-4D)\epsilon_1^3+(22D^2-10D\Delta-5)\epsilon_1^2+(-24D^3+14D^2\Delta+24D-6\Delta)\epsilon_1+9D^4-6\Delta D^3-27D^2+14D\Delta=0$, and $X_{ren}$ and $Z_{ren}$ are functions of the original coupling constants $\Delta$ and $D$ (see Appendix \ref{ExpressionHp}).

\section{hyperdeterminants and Genuine tripartite entanglement}\label{III}
After decomposing the spin chain into blocks and obtaining the renormalization group flow of the coupling constants, some questions naturally arise: what are the tripartite entanglements between the three spin blocks in the low-lying energy eigenstates? How the tripartite entanglement vary as a function of the size of the blocks? And more importantly, can one uses the tripartite entanglement as a probe of the quantum phase transitions? To answer these questions, one needs to first classify and quantify entanglements for tripartite qutrit states. Although multi-partite entanglements are notoriously difficult to quantify due to the fact that almost all polynomial entanglement measures have a degree of at least four \cite{horodecki2009quantum, johansson2014classification}, one may still measure genuine tripartite entanglement by the \textit{hyperdeterminant} \cite{miyake2002multipartite}, which is a generalized determinant for higher-dimensional matrices \cite{gelfand1992hyperdeterminants, gelfand1994discriminants}. Genuine multipartite entangled states are those states with a nonzero hyperdeterminant \cite{miyake2002multipartite}. 

The absolute value of the hyperdeterminant can be used as a multi-partite entanglement measure \cite{miyake2002multipartite}, known as the \textit{concurrence} \cite{hill1997entanglement, wootters1998entanglement} and \textit{three-tangle} \cite{coffman2000distributed} for two- and three-qubit pure states respectively. For general two- and three-qubit states, expressed as $|\Psi^{(2)}\rangle\equiv\sum_{i,j=1,2}\Gamma_{ij}|ij\rangle$ and $|\Psi^{(3)}\rangle\equiv\sum_{i,j,k=1,2}\Gamma_{ijk}|ijk\rangle$, the concurrence and the three-tangle can be explicitly written as $C(|\Psi^{(2)}\rangle)\equiv2|\det(
\mathbf{\Gamma}^{(2)})|=2|\Gamma_{00}\Gamma_{11}-\Gamma_{01}\Gamma_{10}|$ and $\tau_3(|\Psi^{(3)}\rangle)\equiv 4|\Det(\mathbf{\Gamma}^{(3)})|$ respectively \cite{miyake2002multipartite}, where $\mathbf{\Gamma}^{(2)}\equiv[\Gamma_{ij}]$ is a matrix, $\mathbf{\Gamma}^{(3)}\equiv[\Gamma_{ijk}]$ is a third-order tensor, and $\Det(\mathbf{\Gamma}^{(3)})$ is Cayley's hyperdeterminant defined by \cite{miyake2002multipartite, kam2020three}
\begin{align}
    \Det(\mathbf{\Gamma}^{(3)})&=\left(\begin{vmatrix}
    \Gamma_{000}      & \Gamma_{011}\\
    \Gamma_{100}      & \Gamma_{111}
    \end{vmatrix}+
    \begin{vmatrix}
    \Gamma_{010}      & \Gamma_{001}\\
    \Gamma_{110}      & \Gamma_{101}
    \end{vmatrix}\right)^2\nonumber\\
    &-4 \begin{vmatrix}
    \Gamma_{000}      & \Gamma_{001}\\
    \Gamma_{100}      & \Gamma_{101}
    \end{vmatrix}\cdot
     \begin{vmatrix}
    \Gamma_{010}      & \Gamma_{011}\\
    \Gamma_{110}      & \Gamma_{111}
    \end{vmatrix}.
\end{align}

For the present case, one needs to evaluate the hyperdeterminant for tripartite qutrit states. One important properties of the hyperdeterminant is that, under invertible local operations $|\tilde{\psi}\rangle\equiv \mathbf{L}_1\otimes \mathbf{L}_2\otimes\mathbf{L}_3|\psi\rangle$, it transforms with a determinantal factor, $\Det(|\tilde{\psi}\rangle)=\prod_{i=1}^3(\det(\mathbf{L}_i))^2\Det(|\psi\rangle)$, where $\mathbf{L}_i$ ($i=1,2,3$) are invertible $3\times 3$ matrices with complex entries \cite{bengtsson2017geometry}, and $|\psi\rangle$ is an arbitrary tripartite qutrit state. When $\mathbf{L}_i$ are special linear transformations of degree 3, i.e., three-by-three matrices of determinant 1, the hyperdeterminant becomes an invariant under SL$(3,\mathbb{C})^{\otimes 3}$ transformations: $\Det(|\tilde{\psi}\rangle)=\Det(|\psi\rangle)$. Dür showed that if two pure states can be obtained from the other by means of stochastic local operations and classical communications (SLOCC), they have the same kind of entanglement \cite{dur2000three}. They proved that two pure states are equivalent under SLOCC if they can be related by invertible local transformations. In other words, the hyperdeterminant is an SLOCC invariant \cite{dur2000three}.

Since Cayley's hyperdeterminant for two- and three-qubit states and three-qutrit states are homogeneous polynomials of degree $2$, $4$ and $36$ respectively, a direct computation of the three-qutrit hyperdeterminant may be rather involved. Nevertheless, one may still express the three-qutrit hyperdeterminant in terms of the three fundamental SL$(3,\mathbb{C})^{\otimes 3}$ invariants $I_6$, $I_9$ and $I_{12}$, which are homogeneous polynomials of degrees $6$, $9$ and $12$ respectively. The relation between the three-qutrit hyperdeterminant $\Delta_{333}$ and the three fundamental invariants $I_6$, $I_9$ and $I_{12}$ is given by \cite{bremner20143}
\begin{equation}\label{Delta333}
    \Delta_{333}=I_6^3I_9^2-I_6^2J_{12}^2+36I_6I_9^2J_{12}+108I_9^4-32J_{12}^3,
\end{equation}
where $J_{12}\equiv -\frac{1}{24}(I_{12}+I_6^2)$, and the three invariants $I_6$, $I_9$ and $I_{12}$ can be calculated by Cayley's $\Omega$ process (see Appendix \ref{I6I9I12}). After obtaining the hyperdeterminant $\Delta_{333}$, genuine tripartite qutrit entangled states $|\psi\rangle$ are determined by $\Delta_{333}(|\psi\rangle)\neq 0$.

\section{Genuine tripartite entanglement in quantum phase transitions}\label{IV}
With the tripartite qutrit hyperdeterminant $\Delta_{333}$ and the renormalization group flow of the coupling constants, Eq.\:\eqref{RGflow} in hands, one can determine the the tripartite entanglements between the three spin blocks in the low-lying energy eigenstates when the size of the blocks increase, and use the hyperdeterminant as a probe of the quantum phase transitions. In particular, for the case of a large single-ion anisotropy, the lowest energy state $|\psi_0\rangle$ of an isolated spin block can be written as (see Appendix \ref{ExpressionHp})
\begin{align}
    |\psi_0\rangle&= \mathcal{N}_0[|+0-\rangle+|-0+\rangle+a(|+-0\rangle+|-+0\rangle\nonumber\\
    &+|0+-\rangle+|0-+\rangle)+b|000\rangle],
\end{align}
where $a=\epsilon_0/2-D$ and $b=2(1-2D/\epsilon_0)$ are functions of the easy-axis anisotropy parameter $\Delta$ and the single-ion anisotropy parameter $D$, and $\mathcal{N}_0\equiv (2+4a^2+b^2)^{1/2}$ is a normalization constant. A direct computation shows that the three fundamental SL$(3,\mathbb{C})^{\otimes 3}$ invariants $I_6$, $I_9$ and $I_{12}$ for $|\psi_0\rangle$ are (see Appendix \ref{ExpressionHp})
\begin{equation}
    I_6 = -8\mathcal{N}_0^6a^4=\frac{-8a^4}{(2+4a^2+b^2)^3}, I_9=I_{12}=0.
\end{equation}
According to Eq.\:\eqref{Delta333}, it leads to a nonzero three-qutrit hyperdeterminant $\Delta_{333}(|\psi_0\rangle)=\frac{1}{1728}I_6^6$ as long as $a\neq 0$. As the hyperdeterminant $\Delta_{333}(|\psi_0\rangle)$ is proportional to the sixth power of $I_6$, one may use the absolute value of the fundamental invariant $I_6$ to quantify tripartite entanglements between the three spin blocks in the low-lying energy state $|\psi_0\rangle$.

\begin{figure}[tbp]
\includegraphics[width=0.9\columnwidth]{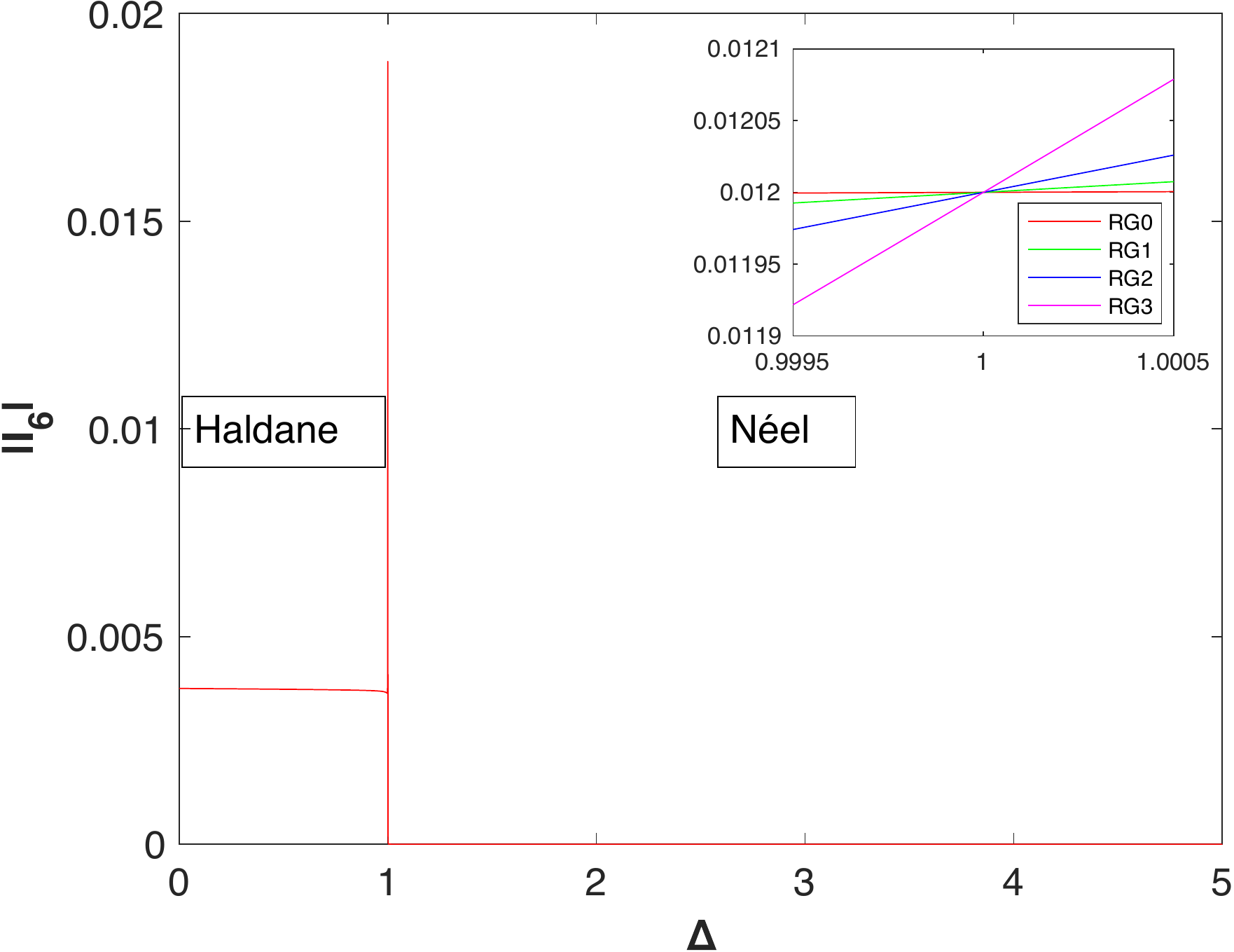}
\caption{Schematic of the tripartite entanglement between the three spin blocks of a Heisenberg spin-1 chain with single-ion anisotropy. Here, we plot the absolute value of the fundamental SL$(3,\mathbb{C})^{\otimes 3}$ invariant $I_6$ for $D=0$ after the ninth step of renormalization, which corresponds to a spin chain of size $N=3^{10}$ with three spin blocks of size $3^9$. The quantum phase transition point which separates the N\'eel and Haldane phases is determined by the crossing of $|I_6|$ for different steps of renormalization.}
\label{D0}
\end{figure}

In Fig.\:\ref{D0}, we plot the absolute value of the fundamental invariant $I_6$ for the low-lying energy state $|\psi_0\rangle$ as a function of $\Delta$ with $D=0$. The result shows that the are two different phases, namely the topological Haldane phase, characterized by a nonzero plateau in $|I_6|$, and the antiferromagnetic N\'eel phase, characterized by a vanishing $|I_6|$. The Ising transition between the N\'eel and Haldane phases at $\Delta=1$ is characterized by a sharp peak in $|I_6|$. It implies that there exists nonzero tripartite entanglement between the three spin blocks in the Haldane phase in the low-lying energy state $|\psi_0\rangle$, while on the contrary, there is no genuine tripartite entanglement between the three spin blocks in the N\'eel phase in the low-lying energy state $|\psi_0\rangle$.

\begin{figure}[tbp]
\includegraphics[width=0.9\columnwidth]{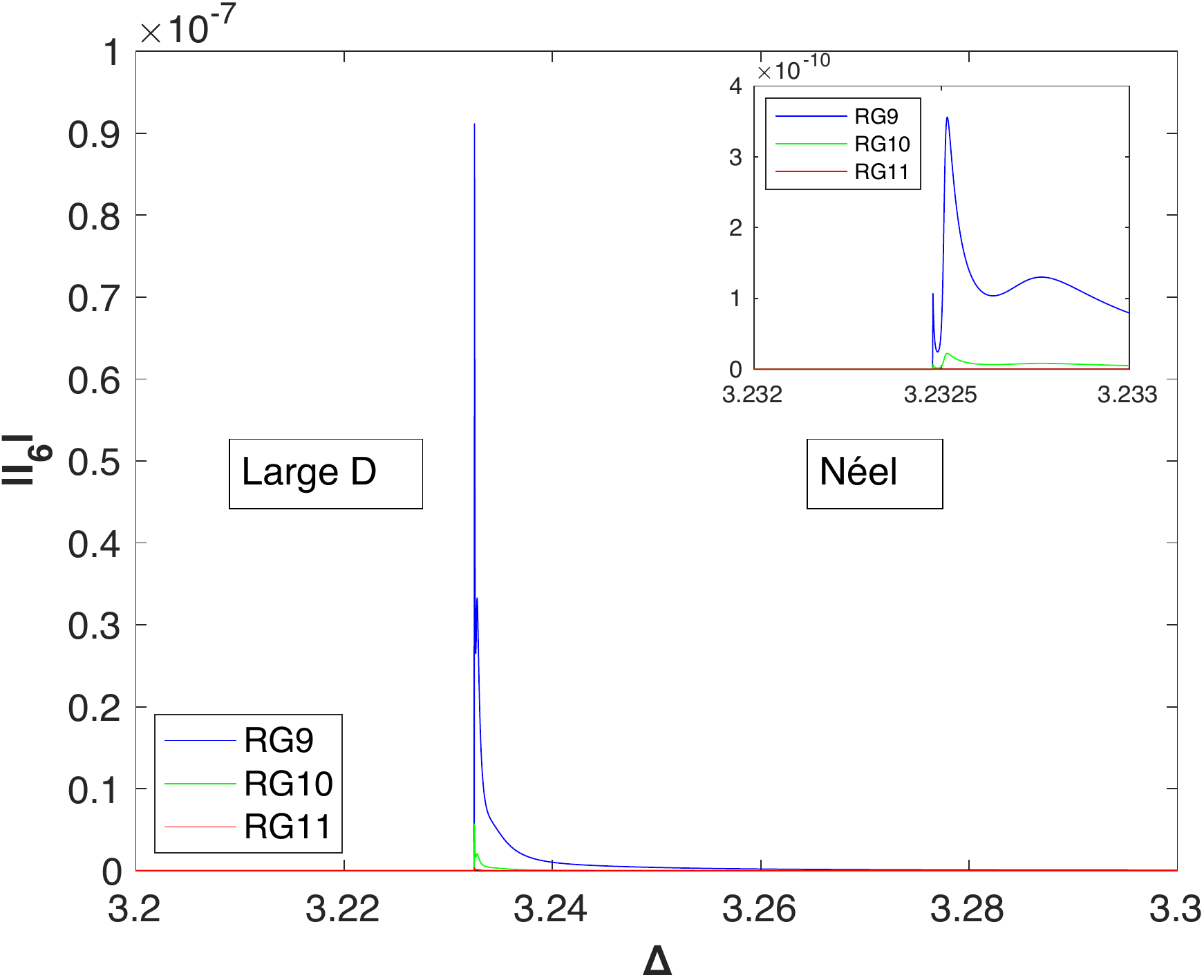}
\caption{Schematic of the tripartite entanglement between the three spin blocks of a Heisenberg spin-1 chain with single-ion anisotropy. Here, we plot the absolute value of the fundamental SL$(3,\mathbb{C})^{\otimes 3}$ invariant $I_6$ for $D=2.5$ after the ninth, tenth and eleventh steps of renormalization, which corresponds to a spin chain of size $N=3^n$ and three spin blocks of size $3^{n-1}$ with $n=10$, $11$ and $12$ respectively. The quantum phase transition point which separates the N\'eel and large-D phases is determined by the crossing of $|I_6|$ for different steps of renormalization.}
\label{D2_5}
\end{figure}

Similarly, in Fig.\:\ref{D2_5}, we plot the absolute value of the fundamental invariant $I_6$ as a function of $\Delta$ for the low-lying energy state $|\psi_0\rangle$ with $D=2.5$. The result shows that the are two different phases, namely the large-D and the antiferromagnetic N\'eel phases, both of which are characterized by a vanishing $|I_6|$. The quantum phase transition between the N\'eel and large-D phases at $\Delta\approx 3.2325$ is characterized by a sharp peak in $|I_6|$, which diminishes quickly when the number of steps of renormalization increases. It implies that, unlike the topological Haldane phase, there is no genuine tripartite entanglement between the three spin blocks in both the N\'eel and the large-D phases.

\begin{figure}[tbp]
\includegraphics[width=0.9\columnwidth]{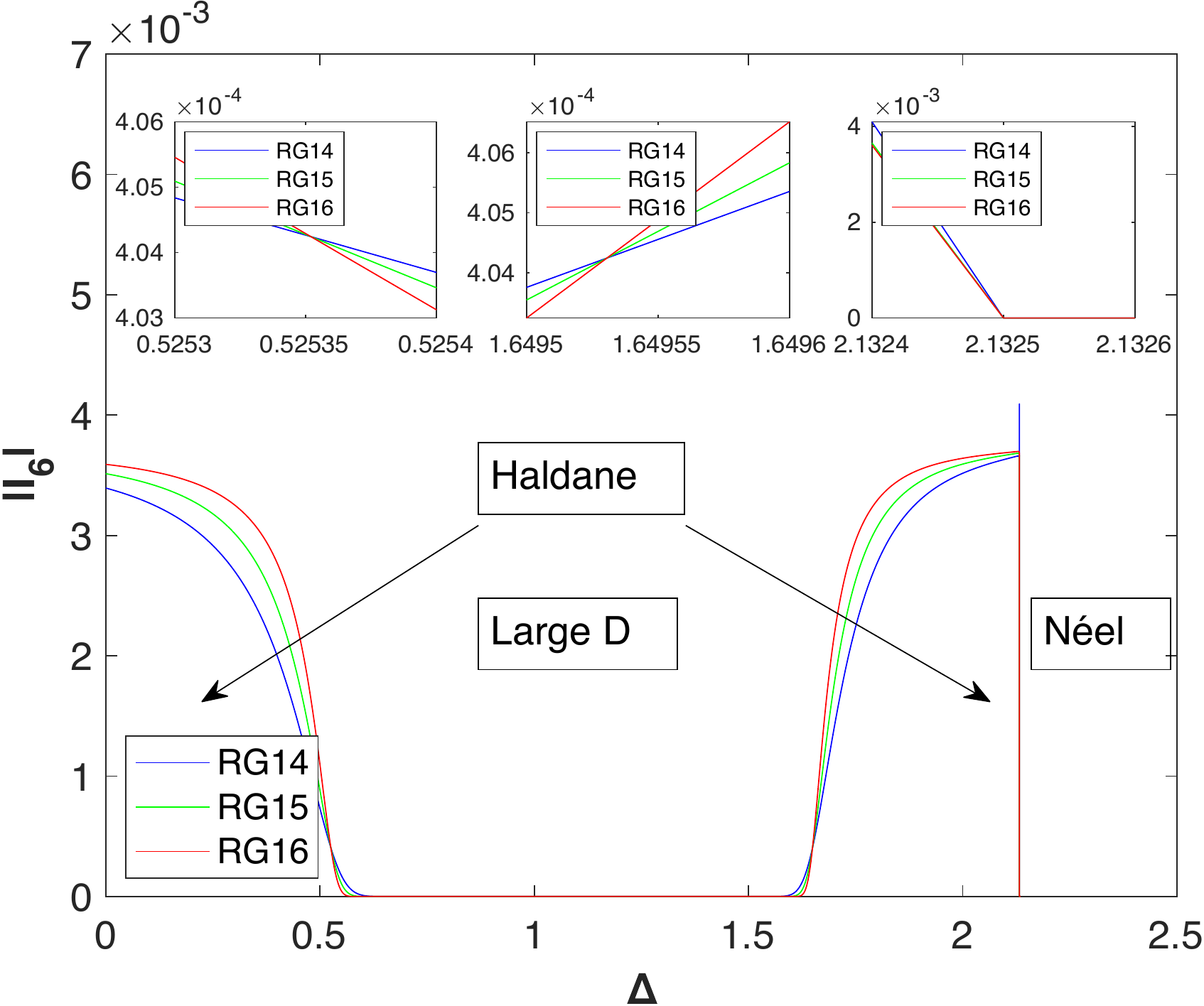}
\caption{Schematic of the tripartite entanglement between the three spin blocks of a Heisenberg spin-1 chain with single-ion anisotropy. Here, we plot the absolute value of the fundamental SL$(3,\mathbb{C})^{\otimes 3}$ invariant $I_6$ for $D=1.4$ after the fourteenth, fifteenth and sixteenth steps of renormalization, which corresponds to a spin chain of size $N=3^n$ and three spin blocks of size $3^{n-1}$ with $n=15$, $16$ and $17$ respectively. The quantum phase transition points which separates the N\'eel, Haldane and large-D phases are determined by the crossing of $|I_6|$ for different steps of renormalization.}
\label{D1_4}
\end{figure}

Finally, in Fig.\:\ref{D1_4}, we plot $|I_6|$ as a function of $\Delta$ for the low-lying energy state $|\psi_0\rangle$ with $D=1.4$. The result shows that there are three different phases, namely the Haldane, the large-D and the antiferromagnetic N\'eel phases. The topological Haldane phases appearing at $0<\Delta<0.52535$ and $1.6495<\Delta<2.1325$ are characterized by two nonzero plateaus in $|I_6|$ with the same height. On the contrary, both of large-D and the antiferromagnetic N\'eel phases are characterized by a vanishing $|I_6|$. The  Gaussian topological phase transition between the  gapful Haldane and large-D phases at $\Delta\approx 0.52535$ and $\Delta\approx 1.6495$ are characterized by an abrupt drop in $|I_6|$, and the Ising transition between the Haldane and N\'eel phases is characterized by a sharp peak in $|I_6|$. It clearly shows that, among the Haldane, large-D and N\'eel phases, only the topological Haldane phase possesses genuine tripartite entanglement between the three spin blocks in the low-lying energy state $|\psi_0\rangle$.

\section{Conclusion}\label{V}
In this work, we set up a quantum information and multipartite entanglement approach to quantum phase transitions in spin chains. In particular, we study the spin-1 Heisenberg XXZ antiferromagnet with a uniaxial single-ion anisotropy term via the quantum renormalization group (QRG) method. We determine the renormalization group flow of the coupling constants, and use the resulting data to plot the tripartite entanglement between the three spin blocks in the low-lying energy eigenstates as a function of the easy-axis anisotropy. We show that topological Haldane phase is characterized by a nonzero plateau in the genuine tripartite entanglement between the spin blocks, while the large-D and N\'eel phases are characterized by a vanishing genuine tripartite entanglement between the spin blocks. Both the N\'eel-Haldane and the N\'eel-large-D phase transitions are distinguished by a sharp peak in the genuine tripartite entanglement, and the Gaussian topological phase transition between the gapful Haldane and large-D phases is distinguished by a quick drop in the genuine tripartite entanglement. Our work reveals that the multipartite entanglement in the low-lying energy eigenstates of spin chains can be used as a probe of quantum phase transitions.

Although our current work is limited to the quantum renormalization group (QRG) method, the current approach can be applied to other numerical methods, such as the density matrix renormalization group method or the tensor renormalization group method. However, due to the fact that nearly all polynomial entanglement measures have a large degree, the calculation of multipartite entanglement in higher spin systems will be more involved.

\begin{appendix}
\section{The explicit expression for the renormalized spin-1 Hamiltonian}\label{ExpressionHp}
In this appendix, we will derive the explicit expression for the renormalized Hamiltonian of a spin-1 Heisenberg chain with single-ion anisotropy. To begin with, one notices that the Hamiltonian for the spin-1 Heisenberg chain with three sites can be explicitly written as
\begin{align}
    H&=J\{(S_x\otimes S_x\otimes I+I\otimes S_x\otimes S_x\nonumber\\
    &+S_y\otimes S_y\otimes I+I\otimes S_y\otimes S_y)\nonumber\\
    &+\Delta(S_z\otimes S_z\otimes I+I\otimes S_z\otimes S_z)\nonumber\\
    &+D(S_z^2\otimes I\otimes I+I\otimes S_z^2\otimes I+I\otimes I\otimes S_z^2)\},
\end{align}
where $J>0$ is the anti-ferromagnetic exchange coupling, $\Delta$ is the easy axis anisotropy, $D$ is the single-ion anisotropy parameter, and $S_x$, $S_y$ and $S_z$ are the spin-1 matrices defined by
\begin{equation}
    S_x\equiv\frac{1}{\sqrt{2}}\begin{pmatrix}
    0 & 1 & 0\\
    1 & 0 & 1 \\
    0 & 1 & 0
\end{pmatrix},
    S_y\equiv\frac{1}{\sqrt{2}}\begin{pmatrix}
    0 & -i & 0\\
    i & 0 & -i \\
    0 & i & 0
\end{pmatrix},
    S_z\equiv\begin{pmatrix}
    1 & 0 & 0\\
    0 & 0 & 0 \\
    0 & 0 & -1
\end{pmatrix}.
\end{equation}

Let us consider the lowest energy eigenstate of the Hamiltonian for a large single-ion anisotropy, which may be expressed as
\begin{align}\label{Psi0}
    |\psi_0&\rangle \equiv \mathcal{N}_0[|+0-\rangle+|-0+\rangle+a(|+-0\rangle+|-+0\rangle\nonumber\\
    &+|0+-\rangle+|0-+\rangle)+b|000\rangle],
\end{align}
where $\mathcal{N}_0\equiv(2+4a^2+b^2)^{-1/2}$ is a normalization constant, $a$ and $b$ are functions of the easy axis anisotropy parameter $\Delta$ and the single-ion anisotropy parameter $D$, and $|+\rangle$, $|0\rangle$ and $|-\rangle$ are the basic kets for the renormalized Hilbert space of each spin block. A direct computation yields
\begin{align}\label{HPsi0}
    H|\psi_0\rangle &=\mathcal{N}_0J\{ 2(D+a)(|+0-\rangle+|-0+\rangle)+4a|000\rangle\nonumber\\
    &+[1+b+a(2D-\Delta)]\nonumber\\
    &\cdot(|+-0\rangle+|-+0\rangle+|0+-\rangle+|0-+\rangle)\}.
\end{align}
Substitution of Eqs.\:\eqref{Psi0} and \eqref{HPsi0} into the eigen-equation $H|\psi_0\rangle=E_0|\psi_0\rangle$ immediately yields $2J(D+a)=E_0$, $J(1+b+a(2D-\Delta))=E_0a$, and $4aJ=E_0b$,
or equivalently
\begin{equation}
a=\frac{\epsilon_0}{2}-D,b=2(1-\frac{2D}{\epsilon_0}),
\end{equation}
where $\epsilon_0\equiv E_0/J$ is the smallest root of the cubic equation
\begin{equation}
    \epsilon_0^3+(\Delta-4D)\epsilon_0^2+(4D^2-2D\Delta-6)\epsilon_0+8D=0.
\end{equation}
For a small single-ion anisotropy, the lowest energy states of the Hamiltonian are doubly degenerate, and are found to be
\begin{subequations}
\begin{align}\label{PsiPlus}
    |\psi_+\rangle &\equiv \mathcal{N}_1[|++-\rangle + |-++\rangle + c(|+00\rangle + |00+\rangle)\nonumber\\
    &+ d|+-+\rangle + e|0+0\rangle],\\
    |\psi_-\rangle &\equiv \mathcal{N}_1[ |--+\rangle + |+--\rangle + c(|-00\rangle + |00-\rangle)\nonumber\\ 
    &+ d|-+-\rangle + e|0-0\rangle],
\end{align}
\end{subequations}
where $\mathcal{N}_1\equiv (2+2c^2+d^2+e^2)^{-1/2}$ is a normalization constant, $c$, $d$ and $e$ are functions of the easy axis anisotropy parameter $\Delta$ and the single-ion anisotropy parameter $D$. A direct computation yields
\begin{align}\label{HPsiPlus}
    H|\psi_+\rangle &= \mathcal{N}_1J\{(c+3D)(|++-\rangle+|-++\rangle)\nonumber\\
    &+[2(c-d\Delta)+3dD]|+-+\rangle+(2c+eD)|0+0\rangle\nonumber\\
    &+(1+cD+d+e)(|+00\rangle+|00+\rangle)\}.
\end{align}
Substitution of Eqs.\:\eqref{PsiPlus} and \eqref{HPsiPlus} into the eigen-equation $H|\psi_+\rangle = E_1|\psi_+\rangle$ immediately yields $J(c+3D)=E_1$, $J(1+cD+d+e)=E_1c$, $J(2c-2d\Delta+3dD)=E_1d$, and $J(2c+eD)=E_1e$,
or equivalently
\begin{equation}
    c=\epsilon_1-3D,d=\frac{2(\epsilon_1-3D)}{\epsilon_1+2\Delta-3D},e=\frac{2(\epsilon_1-3D)}{\epsilon_1-D},
\end{equation}
where $\epsilon_1\equiv E_1/J$ is the smallest root of the quartic equation
\begin{align}
    \epsilon_1^4&+(2\Delta-8D)\epsilon_1^3+(22D^2-10D\Delta-5)\epsilon_1^2\nonumber\\&+(-24D^3+14D^2\Delta+24D-6\Delta)\epsilon_1\nonumber\\&+9D^4-6\Delta D^3-27D^2+14D\Delta=0.
\end{align}

By using the embedding operator, the renormalized Hamiltonian is determined by
\begin{equation}
    H^\prime=\sum_{I=1}^{N/3}(T_I^\dagger h_I^BT_I+T_I^\dagger T_{I+1}^\dagger h_{I,I+1}^{BB}T_{I+1}T_I),
\end{equation}
where the embedding operator for each spin block is constructed by $T_I\equiv |\psi_+\rangle\langle+1|+|\psi_0\rangle\langle 0|+|\psi_-\rangle\langle -1|$, and the Hamiltonians for the isolated spin block, and the inter-block Hamiltonians are described by
\begin{subequations}
\begin{align}
    h_I^B&=J\left[\sum_{j=1}^2(S_{I,j}^xS_{I,j+1}^x+S_{I,j}^yS_{I,j+1}^y+\Delta S_{I,j}^zS_{I,j+1}^z)\right.\nonumber\\
    &\left.+D\sum_{j=1}^3(S_{I,j}^z)^2\right],\\
    h_{I,I+1}^{BB}&=J(S_{I,3}^xS_{I+1,1}^x+J_{I,3}^yS_{I+1,1}^y+\Delta S_{I,3}^zS_{I+1,1}^z).
\end{align}
\end{subequations}
From the eigen-equations $h_I^B|\psi_\pm\rangle=E_1|\psi_\pm\rangle$ and $h_I^B|\psi_0\rangle=E_0|\psi_0\rangle$, one immediately obtains the renormalized Hamiltonian for each isolated spin block
\begin{equation}
    T_I^\dagger h_I^B T_I=E_0I+(E_1-E_0)(S_I^z)^2.
\end{equation}
A direct computation shows that the renormalized spin operators for each spin block are given by
\begin{subequations}
\begin{align}
    T_I^\dagger S_{I,j}^x T_I&= X_{ren}S_I^x,\\
    T_I^\dagger S_{I,j}^y T_I&= X_{ren}S_I^y,\\
    T_I^\dagger S_{I,j}^z T_I &= Z_{ren}S_I^z,
\end{align}
\end{subequations}
where $j=1,3$, $X_{ren} = \mathcal{N}_0\mathcal{N}_1(a+c+bc+ad+ae)$, and $Z_{ren}=\mathcal{N}_1^2(c^2 + d^2)$.
Hence, one obtains the renormalized Hamiltonian for the inter-block Hamiltonian
\begin{align}
    T_I^\dagger T_{I+1}^\dagger h_{I,I+1}^{BB}T_{I+1}T_I&=J[X_{ren}^2(S_I^xS_{I+1}^x+S_I^yS_{I+1}^y)\nonumber\\
    &+\Delta Z_{ren}^2 S_I^zS_{I+1}^z],
\end{align}
which yields the full renormalized Hamiltonian for a spin-1 Heisenberg chain with single-ion anisotropy
\begin{align}\label{HPrime}
    H^\prime =\sum_{I=1}^{N/3}J^\prime[S_I^xS_{I+1}^x+S_I^yS_{I+1}^y+\Delta^\prime S_I^zS_{I+1}^z+D^\prime(S_I^z)^2],
\end{align}
where the constant term $E_0I$ is removed from Eq.\:\eqref{HPrime}, and the renormalized anti-ferromagnetic exchange coupling, easy axis anisotropy, and single-ion anisotropy parameter are given by
\begin{equation}
    J^\prime = X_{ren}^2J,
    \Delta^\prime = \frac{Z_{ren}^2}{X_{ren}^2}\Delta,D^\prime = \frac{\epsilon_1-\epsilon_0}{X_{ren}^2}.
\end{equation}

 \section{The fundamental SL$(3,\mathbb{C})^{\otimes 3}$ invariants $I_6$, $I_9$, $I_{12}$ and the hyperdeterminant $\Delta_{333}$ for tripartite qutrit states}\label{I6I9I12}
Here we compute the three fundamental SL$(3,\mathbb{C})^{\otimes 3}$ invariants $I_6$, $I_9$ and $I_{12}$ (homogeneous polynomials of degrees 6, 9 and 12) for a general tripartite qutrit state $|\psi\rangle\equiv \sum_{i,j,k=1,2,3}\Gamma_{ijk}|ijk\rangle$ by using Cayley's $\Omega$ process \cite{turnbull1960theory, briand2004moduli}. The process starts by identifying a tripartite qutrit state with a trilinear form $f(\mathbf{x},\mathbf{y},\mathbf{z})\equiv \sum_{i,j,k=1,2,3}\Gamma_{ijk}x_iy_jz_k$. 

As an example, one may consider Nurmiev's normal form of an arbitrary trilinear form of $f(\mathbf{x},\mathbf{y},\mathbf{z})$, which can be explicitly written as \cite{nurmiev2000orbits}
\begin{align}
    f^\prime(\mathbf{x},\mathbf{y},\mathbf{z})&\equiv a_1(x_1y_1z_1+x_2y_2z_2+x_3y_3z_3)\nonumber\\
    &+a_2(x_1y_2z_3+x_2y_3z_1+x_3y_1z_2)\nonumber\\
    &+a_3(x_1y_3z_2+x_2y_1z_3+x_3y_2z_1).
\end{align}
As a result of the classical invariant theory \cite{olver1999classical}, the first non-zero fundamental invariant of degree 6 can be constructed via Cayley's $\Omega$ process as \cite{briand2004moduli}
\begin{equation}
    I_6\equiv\frac{1}{1152}\tr\Omega_x^2\Omega_y^2\Omega_z^2\prod_{i=1}^3f^2(\mathbf{x}^{(i)},\mathbf{y}^{(i)},\mathbf{z}^{(i)}),
\end{equation}
where $\tr$ is Olver's trace notation defined by \cite{olver1999classical}
\begin{equation}
    \tr F_1(\mathbf{x}^{(1)})F_2(\mathbf{x}^{(2)})F_3(\mathbf{x}^{(3)})=F_1(\mathbf{x})F_2(\mathbf{x})F_3(\mathbf{x}),
\end{equation}
$\mathbf{x}\equiv (x_1,x_2,x_3)$ is a ternary variable, and $\Omega_x$, $\Omega_y$ and $\Omega_z$ are Cayley's differential operators defined by \cite{briand2004moduli}
\begin{equation}
    \Omega_x\equiv\det(\partial x_i^{(j)}),\:
    \Omega_y\equiv\det(\partial y_i^{(j)}),\:
    \Omega_z\equiv\det(\partial z_i^{(j)}).
\end{equation}
A direct computation yields the fundamental invariant $I_6$ for Nurmiev's normal form 
\begin{equation}
I_6=a_1^6 + a_2^6 + a_3^6 - 10a_1^3a_2^3 - 10a_1^3a_3^3  - 10a_2^3a_3^3.
\end{equation}
Similarly, the last non-zero fundamental invariant of degree 12 can be constructed by Cayley's $\Omega$ process as \cite{briand2004moduli}
\begin{subequations}
\begin{align}
    I_{12}&\equiv \frac{1}{124416}\tr\Omega_x^4\Omega_y\Omega_z\prod_{i=1}^3B_\alpha(\mathbf{x}^{(i)})f(\mathbf{x}^{(i)},\mathbf{y}^{(i)},\mathbf{z}^{(i)}),\\
    B_\alpha&\equiv \tr\Omega_y\Omega_z\prod_{i=1}^3f(\mathbf{x}^{(i)},\mathbf{y}^{(i)},\mathbf{z}^{(i)}),
\end{align}
\end{subequations}
where $B_\alpha$ is an auxiliary  degree-6 homogeneous polynomial of the coefficients $\Gamma_{ijk}$ and the variables $x_1$, $x_2$  and $x_3$. A direct computation yields the fundamental invariant $I_{12}$ and the auxiliary polynomial $B_\alpha$ for Nurmiev's normal form
\begin{subequations}
\begin{align}
    I_{12}&=-\mu\left(\mu^3+(6\nu)^3\right),\\
    B_{\alpha}&=6\left(\mu x_1x_2x_3 - \nu(x_1^3+x_2^3+x_3^3)\right),
\end{align}
\end{subequations}
where $\mu\equiv a_1^3+a_2^3+a_3^3$ and $\nu\equiv a_1a_2a_3$.

Notice that although the fundamental invariants $I_6$ and $I_{9}$ are determined up to non-zero scalar multiples, the fundamental invariant $I_{12}$ is \textit{only} determined up to a scalar multiple of $I_6^2$ \cite{bremner20143}. As a result,  there are different conventions for the fundamental invariant of degree 12. A direct computation shows that Bremner's invariant (denoted as $J_{12}$) and Briand's invariant (denoted as $I_{12}$) satisfy the simple relation $-I_{12}-I_6^2 = 24J_{12}$. Hence, Bremner's invariant $J_{12}$ for Nurmiev’s normal form can be explicitly written as \cite{bremner20143}
\begin{align}
    J_{12} &= a_1^3a_2^9+
    a_1^9a_2^3 +
    a_1^3a_3^9 +
    a_1^9a_3^3 +  a_2^3a_3^9 + a_2^9a_3^3 +2a_1^3a_2^3a_3^6   \nonumber\\
    &+ 2a_1^3a_2^6a_3^3 + 2a_1^6a_2^3a_3^3 - 4a_1^6a_2^6  - 4a_1^6a_3^6  - 4a_2^6a_3^6.
\end{align}
Finally, the second non-zero fundamental invariant of degree 9 can be constructed by Cayley's $\Omega$ process as \cite{briand2004moduli}
\begin{equation}
I_9\equiv\frac{1}{576}\tr \Omega_x\Omega_y\Omega_z\Omega_{\xi}\Omega_{\eta}\Omega_{\zeta}E_{\alpha}^{(1)}E_{\beta}^{(2)}E_{\beta}^{(3)},
\end{equation}
where $\xi$, $\eta$ and $\zeta$ are some auxiliary variables, and $E_{\alpha}^{(i)}$ and $E_{\beta}^{(i)}$ are short notations for $E_{\alpha}(\mathbf{x}^{(i)},\mathbf{y}^{(i)},\mathbf{z}^{(i)},\boldsymbol{\xi}^{(i)},\boldsymbol{\eta}^{(i)},\boldsymbol{\zeta}^{(i)})$ and $E_{\beta}(\mathbf{x}^{(i)},\mathbf{y}^{(i)},\mathbf{z}^{(i)},\boldsymbol{\xi}^{(i)},\boldsymbol{\eta}^{(i)},\boldsymbol{\zeta}^{(i)})$, which are defined by
\begin{subequations}
\begin{gather}
    E_{\alpha}\equiv \tr\Omega_x Q_{\alpha}^{(1)}f^{(2)}P_{\alpha}^{(3)}\:\:\mbox{and}\:\:E_{\beta}\equiv \tr\Omega_y Q_{\beta}^{(1)}f^{(2)}P_{\beta}^{(3)}.\\
    Q_{\alpha}\equiv \tr\Omega_y\Omega_z f^{(1)}f^{(2)}P_{\beta}^{(3)}P_{\gamma}^{(3)},\\
    Q_{\beta}\equiv \tr\Omega_x\Omega_z f^{(1)}f^{(2)}P_{\alpha}^{(3)}P_{\gamma}^{(3)}.
\end{gather}
\end{subequations}
Here, $Q_{\alpha}^{(1)}$ and $Q_{\beta}^{(1)}$ are short notations for $Q_{\alpha}(\mathbf{x}^{(1)},\boldsymbol{\eta}^{(1)},\boldsymbol{\xi}^{(1)})$ and $Q_{\beta}(\boldsymbol{\xi}^{(1)},\mathbf{y}^{(1)},\boldsymbol{\eta}^{(1)})$, $f^{(i)}\equiv f(\mathbf{x}^{(i)},\mathbf{y}^{(i)},\mathbf{z}^{(i)})$, $P_\alpha^{(3)}\equiv (\mathbf{x}^{(3)}\cdot\boldsymbol{\xi}^{(3)})$, $P_\beta^{(3)}\equiv (\mathbf{y}^{(3)}\cdot\boldsymbol{\eta}^{(3)})$, $P_\gamma^{(3)}\equiv (\mathbf{z}^{(3)}\cdot\boldsymbol{\zeta}^{(3)})$, and  $\boldsymbol{\xi}\equiv(\xi_1,\xi_2,\xi_3)$ and $\boldsymbol{\eta}\equiv(\eta_1,\eta_2,\eta_3)$ are the auxiliary ternary variables.

A direct computation yields $Q_{\alpha}$, $Q_{\beta}$, $E_{\alpha}$, $E_{\beta}$ and the fundamental invariant $I_9$ for Nurmiev's normal form \cite{bremner20143}
\begin{subequations}
\begin{align}
    I_9&=-(a_1^3-a_2^3)(a_1^3-a_3^3)(a_2^3-a_3^3),\\
    Q_{\alpha}&=2\sum_{ijk}x_ix_j(a_1^2\eta_k\zeta_k+a_2^2\eta_i\zeta_j+a_3^2\eta_j\zeta_i)\nonumber\\
    &- 2\sum_{ijk}x_i^2(a_1a_2\eta_k\zeta_j+a_1a_3\eta_j\zeta_k+a_2a_3\eta_i\zeta_i),\\
    E_{\alpha}&=2\sum_{ijk}\{\xi_i(a_1y_kz_k+a_2y_iz_j + a_3y_jz_i)[a_1^2(x_k\eta_i\zeta_i+x_i\eta_k\zeta_k)   \nonumber\\
    &+a_2^2(x_i\eta_i\zeta_j+ x_k\eta_j\zeta_k) +a_3^2(x_i\eta_j\zeta_i + x_k\eta_k\zeta_j) \nonumber\\
    &-2x_j(a_1a_2\eta_i\zeta_k + a_1a_3\eta_k\zeta_i + a_2a_3\eta_j\zeta_j)] \nonumber\\
    &-\xi_i(a_1y_jz_j + a_2y_kz_i + a_3y_iz_k)[a_1^2(x_j\eta_i\zeta_i + x_i\eta_j\zeta_j) \nonumber\\
    &+a_2^2(x_i\eta_k\zeta_i
    + x_j\eta_j\zeta_k) +a_3^2(x_i\eta_i\zeta_k + x_j\eta_k\zeta_j) \nonumber\\
    &-2x_k(a_1a_2\eta_j\zeta_i + a_1a_3\eta_i\zeta_j + a_2a_3\eta_k\zeta_k)]\},
\end{align}
\end{subequations}
where the summation is performed over all cyclic permutations of $1,2,3$, $Q_{\beta}$ is obtained from $Q_{\alpha}$ by replacing $x_i$, $\eta_i$ and $\zeta_i$ with $y_i$, $\zeta_i$ and $\xi_i$ respectively, and $E_{\beta}$ is obtained from $E_{\alpha}$ by performing the cyclic permutations $(x_i,y_i,z_i)\rightarrow(y_i,z_i,x_i)$ and $(\xi_i,\eta_i,\zeta_i)\rightarrow(\eta_i,\zeta_i,\xi_i)$.

According to Theorem 3.1 of \cite{bremner20143}, the $3\times 3\times 3$ hyperdeterminant $\Delta_{333}$ (homogeneous polynomial of degree 36) for a general trilinear form can be generated from the three fundamental SL$(3,\mathbb{C})^{\otimes 3}$ invariants $I_6$, $I_9$ and $J_{12}$ as 
\begin{equation}
\Delta_{333}=I_6^3I_9^2-I_6^2J_{12}^2+36I_6I_9^2J_{12}+108I_9^4-32J_{12}^3.\label{Hyperdeterminant}
\end{equation}
In particular, for the case of a large single-ion anisotropy, the lowest energy state $|\psi_0\rangle$ of an isolated spin block of a spin-1 Heisenberg chain has the form
\begin{align}
    |\psi_0\rangle&\equiv \mathcal{N}_0[(|321\rangle+|123\rangle)+a(|312\rangle+|132\rangle\nonumber\\
    &+|231\rangle +|213\rangle)+b|222\rangle],
\end{align}
where $\mathcal{N}_0$ is a normalization constant, and $a$ and $b$ are functions of the single-ion anisotropy parameter $D$. Here, we have identified the basic kets for the renormalized
Hilbert space of each spin block, namely $|+\rangle$, $|0\rangle$, and $|-\rangle$, as the qutrit states $|3\rangle$, $|2\rangle$, and $|1\rangle$ respectively. As we have explained before, the state $|\psi_0\rangle$ can be identified with the following trilinear form
\begin{align}
    f_0(\mathbf{x},\mathbf{y},\mathbf{z})&=\mathcal{N}_0[(x_3y_2z_1+x_1y_2z_3)+a(x_3y_1z_2+x_1y_3z_2\nonumber\\
    &+x_2y_3z_1+x_2y_1z_3)+bx_2y_2z_2].
\end{align}
A direct computation yields $B_{\alpha}$, $Q_{\alpha}$, $Q_{\beta}$, $E_{\alpha}$ and $E_{\beta}$ for the trilinear form $f_0(\mathbf{x},\mathbf{y},\mathbf{z})$
\begin{align}
    B_{\alpha}&= 
    6\mathcal{N}_0^3a^2(2x_1x_2x_3-bx_2^3),\label{Invariant1}\\
    Q_{\alpha}&=2\mathcal{N}_0^2a\{[x_1x_2e_{(12)}^\prime+x_1x_3e_{(13)}+x_2x_3e_{(32)}^\prime]\nonumber\\
    &-[(x_1^2e_{11}+x_3^2e_{33})+x_2^2(be_{(13)}+ae_{22})]\},\\
    Q_{\beta}&=2\mathcal{N}_0^2\{a[y_1y_2f_{(12)} +ay_1y_3f_{(13)} + y_2y_3f_{(23)}
     \nonumber\\
    &- a(y_1^2f_{11} +y_3^2f_{33}) ]-y_2^2( f_{22}+bf_{(13)})\},\\
    E_{\alpha}&=2\mathcal{N}_0^2a\{(\xi_2\pi_1-\xi_1\pi_2)(x_1e_{(31)}+ x_2e_{(32)}^\prime -x_3e_{(33)})\nonumber\\
    &+(\xi_1\pi_3-\xi_3\pi_1)[x_1e_{(12)}^\prime+ x_3e_{(32)}^\prime- x_2(ae_{(22)} + 2be_{(13)})]\nonumber\\
    &+ (\xi_3\pi_2-\xi_2\pi_3)(x_3e_{(13)}+x_2e_{(12)}^\prime -x_1e_{(11)})\},\\
    E_{\beta}&=
    2\mathcal{N}_0^2\{a(\eta_2\Pi_1-\eta_1\Pi_2)[a(y_1f_{(31)}- y_3f_{(33)}) + y_2f_{(32)}]\nonumber\\
    &+(\eta_1\Pi_3-\eta_3\Pi_1)[a(y_1f_{(21)} + y_3f_{(23)})- y_2(f_{(22)}+2bf_{(13)})]\nonumber\\
    &+a(\eta_3\Pi_2-a\eta_2\Pi_3)[a(y_3f_{(13)}- y_1f_{(11)})  + y_2f_{(12)}]\}\label{Invariant5},
\end{align}
where $e_{ij}\equiv \eta_i\zeta_j$, $e_{(ij)}\equiv e_{ij}+e_{ji}$, $e_{(ij)}^\prime\equiv e_{ij}+ae_{ji}$, $f_{ij}\equiv \xi_i\zeta_j$, $f_{(ij)}\equiv f_{ij}+f_{ji}$, $\pi_i\equiv \partial f_0/\partial x_i$, and $\Pi_i\equiv \partial f_0/\partial y_i$. Using Eqs.\:\eqref{Invariant1} - \eqref{Invariant5}, we obtain the final expressions of the three fundamental SL$(3,\mathbb{C})^{\otimes 3}$ invariants $I_6$, $I_9$ and $I_{12}$ for $|\psi_0\rangle$:
\begin{equation}
I_6=-8\mathcal{N}_0^6a^4,I_9=I_{12}=0,
\end{equation}
or equivalently
\begin{equation}
    I_6=-8\mathcal{N}_0^6a^4,I_9=0,J_{12}=-\frac{1}{24}I_6^2.\label{Bremner}
\end{equation}
Finally, substitution of Eq.\:\eqref{Bremner} into Eq.\:\eqref{Hyperdeterminant} immediately yields the hyperdeterminant $\Delta_{333}$ for $|\psi_0\rangle$
\begin{equation}
    \Delta_{333}(|\psi_0\rangle)=\frac{1}{1728}I_6^6.
\end{equation}

\end{appendix}
 
\begin{acknowledgements}
The Authors would like to thank the Science and Technology Development Fund of the Macau SAR for providing support, FDCT 023/2017/A1. The Authors would also like to thank the University of Macau in providing sup- port, MYRG2018-00125-FST.
\end{acknowledgements}

\end{document}